\documentclass[%
 preprint,
 amsmath,amssymb,
 aps,
pra,
]{revtex4-1}

\usepackage{graphicx}
\usepackage{dcolumn}
\usepackage{bm}
\usepackage{lipsum}
\usepackage[capitalise]{cleveref}

\newcommand{\kk}{\bm{k}}
\newcommand{\rr}{\bm{r}}

\begin{document}


\title{Electronic Response Quantities of Solids and Deep Learning}

\author{Kevin Ryczko}
\email{kevin.ryczko@uottawa.ca}
\affiliation{Department of Physics, University of Ottawa}
\affiliation{Vector Institute for Artificial Intelligence}
\author{Olivier Malenfant-Thuot}
\affiliation{D\'epartement de Physique, Universit\'e de Montr\'eal}
\affiliation{Regroupement Qu\'eb\'ecois sur les Mat\'eriaux de Pointe}
\author{Isaac Tamblyn}
\email{isaac.itamblyn@uottawa.ca}
\affiliation{Department of Physics, University of Ottawa}
\affiliation{Vector Institute for Artificial Intelligence}
\author{Michel C\^ot\'e}
\affiliation{D\'epartement de Physique, Universit\'e de Montr\'eal}
\affiliation{Regroupement Qu\'eb\'ecois sur les Mat\'eriaux de Pointe}

\date{\today}

\begin{abstract}
We introduce a deep neural network (DNN) framework called the \textbf{r}eal-space \textbf{a}tomic \textbf{d}ecomposition \textbf{net}work (\textsc{radnet}), which is capable of making accurate polarization and static dielectric function predictions for solids. We use these predictions to calculate Born-effective charges, longitudinal optical transverse optical (LO-TO) splitting frequencies, and Raman tensors for two prototypical examples: GaAs and BN. We then compute the Raman spectra, and find excellent agreement with \textit{ab initio} techniques. \textsc{radnet} is as good or better than current methodologies. Lastly, we discuss how \textsc{radnet} scales to larger systems, paving the way for predictions of response functions on meso-scale structures with \textit{ab initio} accuracy. 
\end{abstract}

\pacs{Valid PACS appear here}
\maketitle


\section{Introduction}
\label{intro}
Advances in the application of machine learning in quantum chemistry and condensed matter physics have shown that one can accurately model the potential energy surfaces of molecules and materials \cite{Noe2020, von2020retrospective, mueller2020machine, unke2020machine}. These machine learning models approach the accuracy of the electronic structure approach they were trained with, and demonstrate significant computational speed-ups compared to traditional electronic structure calculations. The next obvious application of machine learning is modeling how these potential energy surfaces change in response to a perturbation. In condensed matter physics, two commonly studied perturbations are atomic displacements and external electric fields. First-order responses of these quantities give forces and polarization. Second-order responses give phonons, electronic susceptibility, and effective charges. Third-order responses include Raman tensors, non-linear electronic susceptibility, and phonon-phonon interactions. A multitude of such quantities have yet to be reliably computed for solids with machine learning in a scalable manner. \\

For isolated molecules, various machine learning techniques have been introduced and applied to predict response quantities \cite{grisafi2018symmetry, wilkins2019accurate, christensen2019operators, gastegger2020machine}. In Ref. \cite{grisafi2018symmetry}, a symmetry-adapted machine learning framework was introduced for the prediction of the dipole moment, polarization, and hyper-polarization of water oligomers. This methodology was also applied to the QM7b database \cite{wilkins2019accurate}, which contains coupled-cluster polarizabilities for various molecules. In Ref. \cite{christensen2019operators}, the operator quantum machine learning (OQML) framework was introduced based on kernel-based machine learning methods. This framework was used to compute forces, dipole moments, normal modes, and infrared spectra for different molecular datasets. In Ref. \cite{gastegger2020machine}, the FieldSchNet architecture was introduced and utilized to calculate dipoles, polarizabilities, infrared spectra, Raman spectra, and nuclear magnetic resonance shifts for various molecules.\\

For solids, however, machine learning techniques have mostly been applied to responses for atomic displacements \cite{babaei2019machine, rohskopf2020fast, mortazavi2020exploring}. The only reports that focus on other response quantities, to date, calculated dielectric tensors of liquid water and ice \cite{grisafi2018symmetry} or molecular crystals \cite{raimbault2019using}. \\

\begin{figure*}[t]
    \centering
    \includegraphics[width=\linewidth]{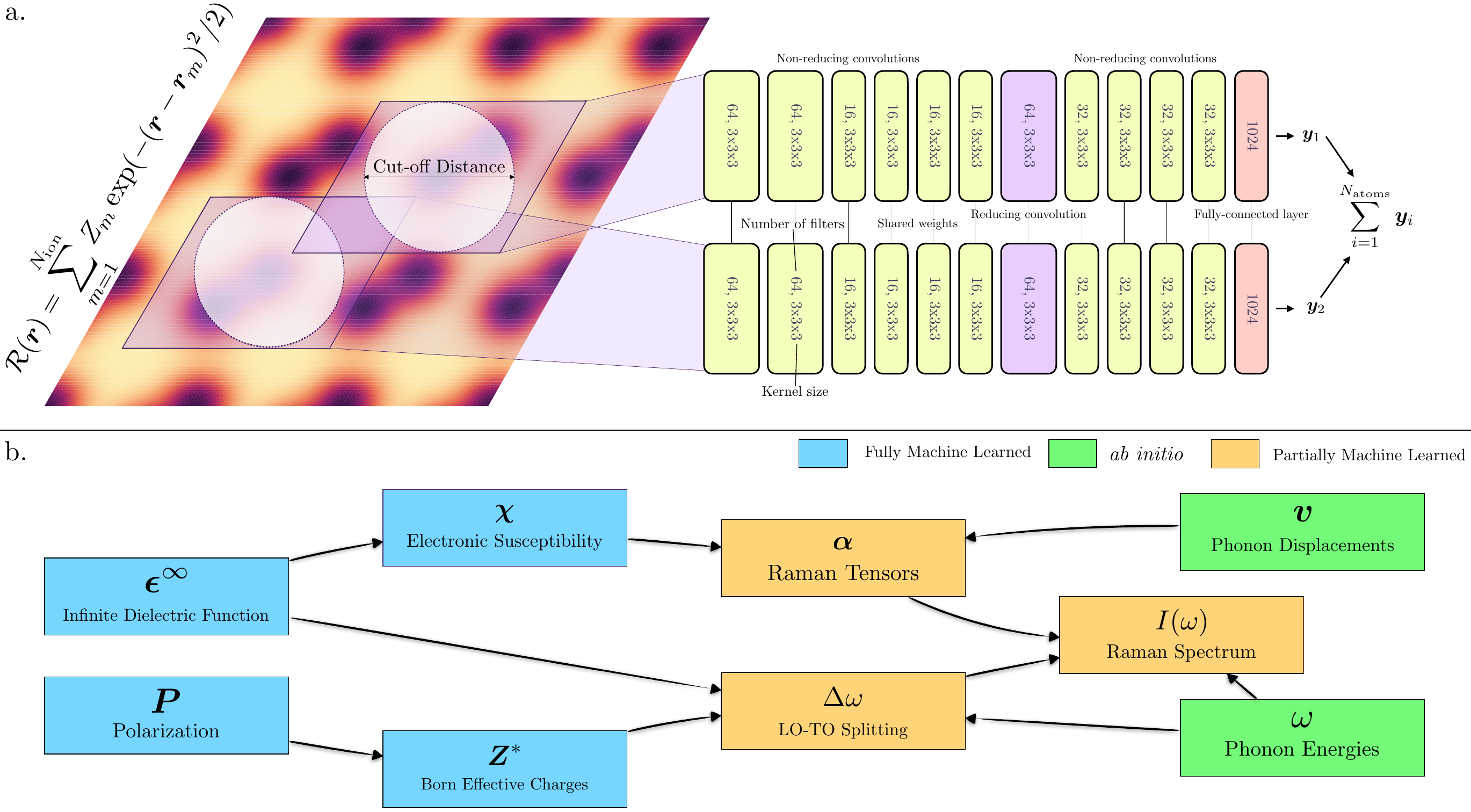}
    \caption{(a) \textsc{radnet}. The system is first represented by a sum over Gaussian functions ($\mathcal{R}(\bm{r})$). 3D slices of $\mathcal{R}(\bm{r})$, centered on the atomic positions are passed through a deep neural network which outputs the atomic contribution to the desired quantity. (b) Quantities computed with either machine learning, \textit{ab initio} techniques, or a hybrid of the two. Lines show the connections between the quantities.}
    \label{fig_1}
\end{figure*}

Here, we introduce a deep learning framework, called the \textbf{r}eal-space \textbf{a}tomic \textbf{d}ecomposition \textbf{net}work (\textsc{radnet}), to predict the polarization and static dielectric tensor. We then use these predictions to calculate Born-effective charges, the longitudinal optical transverse optical (LO-TO) splitting, and Raman tensors. Although machine learning methods have been used to compute the polarization and electronic dielectric function in a condensed matter system \cite{grisafi2018symmetry}, derivatives of these quantities, have yet to be reported. In addition, previous reports focused on phonons for non-metallic systems which \textit{did not} include LO-TO splitting. Therefore, our work allows for a machine-learned correction to LO phonon energies. LO-TO splitting occurs due to induced dipoles in the direction parallel to the applied electric field. In a solid, induced dipoles introduce long-range, dipole-dipole interactions and also produce a macroscopic electric fields. Current work is dedicated to constructing machine learning potentials that include long-range effects \cite{ko2021general}. The outline of this report is as follows: In \cref{el_theory}, we include all of the necessary electronic structure calculations needed to compute the various quantities of interest in this report. In \cref{data_gen}, we cover the dataset generation details. In \cref{deep_learning}, we discuss machine learning methodology and hyper-parameter choices. In \cref{results}, we report and comment on our results and potential future work. Lastly, in \cref{conclusion}, we summarize.

\section{Methods}
\label{methods}

\subsection{Electronic Structure Theory}
\label{el_theory}
We first discuss our \textit{ab initio} computations, beginning with the polarization, $\bm{P}$. The polarization requires derivatives of Kohn-Sham orbitals with respect their wavevectors (which can be calculated via density functional perturbation theory (DFPT) \cite{baroni2001phonons} or Berry-phase theory \cite{king1993theory}). In Berry-phase theory, as demonstrated in the modern theory of polarization \cite{resta2007theory}, the total polarization per unit volume in atomic units is 
\begin{equation}
\label{berry}
    \bm{P} = \frac{e}{(2\pi)^3}\text{Im}\sum_{n=1}^{N_{\text{el}} / 2}\int_{\text{BZ}} d\bm{k}~\langle \psi_{n\bm{k}}|\nabla_{\bm{k}}|\psi_{n\bm{k}}\rangle + \frac{1}{\Omega_0}\sum_{m=1}^{N_\text{ion}}Z_m \bm{r}_m,
\end{equation}
where $n$ is the band index, $N_{\text{el}}$ is the number of electrons, $\psi_{n\bm{k}}$ is a Kohn-Sham orbital, $\Omega_0$ is the unit cell volume, $N_{\text{ion}}$ is the number of ions, and $Z_m$ is the atomic number ion $m$.  For multi-band systems, the electronic contribution is written in terms of overlap matrices between Kohn-Sham orbitals of neighbouring $\bm{k}$-points, as demonstrated in Ref. \cite{sai2002theory}. Once the polarization is known, the Born-effective charge, defined for a particular atom $m$
\begin{equation}
\label{bo_charge}
    Z_{m, \beta\alpha}^* = \Omega_0\frac{\partial P_{\beta}}{\partial \tau_{m, \alpha}(\bm{q}=0)}
\end{equation}
can be computed with DFPT \cite{gonze1997dynamical} or finite differences. In this work, we use finite differences to calculate derivatives of the the machine learning model predictions. In \cref{bo_charge}, $\tau_{m, \alpha}$ is the perturbation to the atomic coordinate of atom $m$ in the direction $\alpha$. 


Next, we discuss the computation of the electric susceptibility, $\bm{\chi}$, the static dielectric tensor, $\bm{\epsilon}^{\infty}$, and the Raman tensor, $\bm{\alpha}$. The electric susceptibility can be written as
\begin{equation}
    \chi_{ij} = \frac{\partial ^2 E }{\partial \mathcal{E}_i \partial \mathcal{E}_j}
\end{equation}
where $E$ is the total energy and $\bm{\mathcal{E}}$ is an electric field. The static electronic dielectric tensor can then be defined as 
\begin{equation}
    \epsilon_{ij}^{\infty} = \delta_{ij} - \frac{8\pi}{\Omega_0}\chi_{ij},
\end{equation}
where $\delta_{ij}$ is the Kronecker delta function. The Raman tensor adds an additional derivative to the energy with respect to an atomic displacement. For each phonon mode $\ell$, the Raman tensor is defined as 
\begin{equation}
\label{raman-tensors}
    \alpha_{ij}^{\ell}= \sum_{\alpha=1}^3\sum_{m=1}^{N_{\text{ion}}}\frac{\partial \chi_{ij}}{\partial \tau_{m, \alpha}} v_{\ell}(m, \alpha)
\end{equation}
where and $v_{\ell}$ is a phonon eigenvector. Derivatives of the electronic susceptibility can also be computed using finite differences or DFPT. Using DFPT, the electronic susceptibility, as shown in \cite{gonze1997dynamical}, is computed via
\begin{equation}
    \chi_{ij}(\rr) = -\text{Im}\frac{\Omega_0}{(2\pi)^3}\sum_{n=1}^{N_{\text{el}}}\int_{\text{BZ}}d\kk~\langle {u_{n\kk}^{\mathcal{E}_i}}|u_{n\kk}^{k_j}\rangle,
\end{equation}
where ${u_{n\kk}^{\mathcal{E}_i}}$ is the first order correction to the ground state Kohn-Sham orbital $u_{n\kk}$ with respect to an electric field $\bm{\mathcal{E}}$ applied in direction $i$, and $u_{n\kk}^{k_j}$ is the periodic part of the Kohn-Sham orbital which has been differentiated with respect to its wavevector $\kk$ in direction $j$.\\

With all the fundamental quantities now defined, we discuss the computation of LO-TO splitting and computation of the Raman spectrum. As shown in \cite{gonze1997dynamical}, the LO and and TO modes split due to an induced dipole in the direction of the applied electric field, and the difference between the squared phonon energies as $\bm{q}\rightarrow0$ is
\begin{equation}
\label{loto}
    \Delta \omega(\bm{q}\rightarrow 0) = \frac{4\pi}{\Omega_0}\sum_{m=1}^{N_{\text{ion}}}\frac{1}{M_m}\frac{\sum_{\alpha\beta\gamma=1}^3q_{\alpha}Z_{m,\alpha\beta}^*Z_{m,\gamma\beta}^*q_{\gamma}}{\sum_{\alpha\beta=1}^3q_{\alpha}\epsilon_{\alpha\beta}^{\infty}q_{\beta}},
\end{equation}
where $M_m$ is the mass of atom $m$. As shown in Ref. \cite{romero2020abinit}, the contribution of the phonon mode $\ell$, to the Raman spectrum is
\begin{equation}
\label{raman}
    I^{\ell}(\omega)=2\pi C^{\ell}(\omega)[(10G_0^{\ell} + 4 G_2^{\ell}) + (5G_1^{\ell} + 3G_2^{\ell})],
\end{equation}
with 
\begin{equation}
\label{freq_raman}
    C^{\ell}(\omega) = \frac{(\omega_{\ell} - \omega_I)^4}{2\omega_{\ell}c^4}[n(\omega_{\ell}) + 1]\frac{\Gamma}{(\omega - \omega_{\ell})^2 + \Gamma^2}.
\end{equation}
In \cref{raman}, the rotation invariants are \cite{caracas2006theoretical}
\begin{eqnarray}
G_0&=&\frac{1}{3}(\alpha_{11}^2 + \alpha_{22}^2 + \alpha_{33}^2),\\
G_1&=&\frac{1}{2}[(\alpha_{12} - \alpha_{21})^2 + (\alpha_{23} - \alpha_{32})^2 + (\alpha_{13} - \alpha_{31})^2],\nonumber\\
\\
G_2&=&\frac{1}{2}[(\alpha_{12} + \alpha_{21})^2 + (\alpha_{23} + \alpha_{32})^2 + (\alpha_{13} + \alpha_{31})^2],\nonumber\\
\end{eqnarray}
and in \cref{freq_raman}, $\omega_{\ell}$ is the frequency of the $\ell$th phonon, $\omega_I$ is the frequency of light, $c$ is the speed of light, $n(\omega)$ is the Bose-Einstein distribution, and $\Gamma$ is a broadening factor to simulate impurities that would be present experimentally. For the Bose-Einstein distribution, we used a temperature of 300 K. For $\omega_I$, we used a value of 532~nm, which has been used in a previous report to study the Raman spectra of MoS${}_2$ \cite{mignuzzi2015effect}. Additionally, we used $\Gamma=10\text{ cm}^{-1}$ for the broadening factor.

\subsection{Data Generation}
\label{data_gen}
To demonstrate our approach, we focus on two example solids: GaAs and cubic BN. We defined 2 atom unit cells for both systems and computed the polarization and infinite dielectric tensor for different random configurations of the atomic positions. For GaAs, the lattice constant was 4.066~\AA~ and for BN the lattice constant was 2.564~\AA.  The calculations were done using the PBE exchange-correlation functional, a $12\times12\times12$, unshifted $\bm{k}$-point grid, and an energy cutoff of 45 Ha. We used Abinit \cite{gonze2020abinit, romero2020abinit, Gonze1997, gonze1997dynamical} for all of our electronic structure calculations. To generate the random configurations, we used normal distributions centered at the equilibrium positions with a standard deviation of 0.01 \AA. The number of random configurations were 1200 for GaAs and 1000 for BN. In the additional 200 configurations for GaAs, the $x$-coordinates of the atoms were held fixed which allowed for an improved prediction of the infinite dielectric tensor. This is described in further detail in \cref{results}. Of the configurations,  20\% were used as a validation set during training. The final tests were made with comparisons of the predicted quantities for the equilibrium structures unseen during training. \\

\begin{figure}[t]
    \centering
    \includegraphics[width=\linewidth]{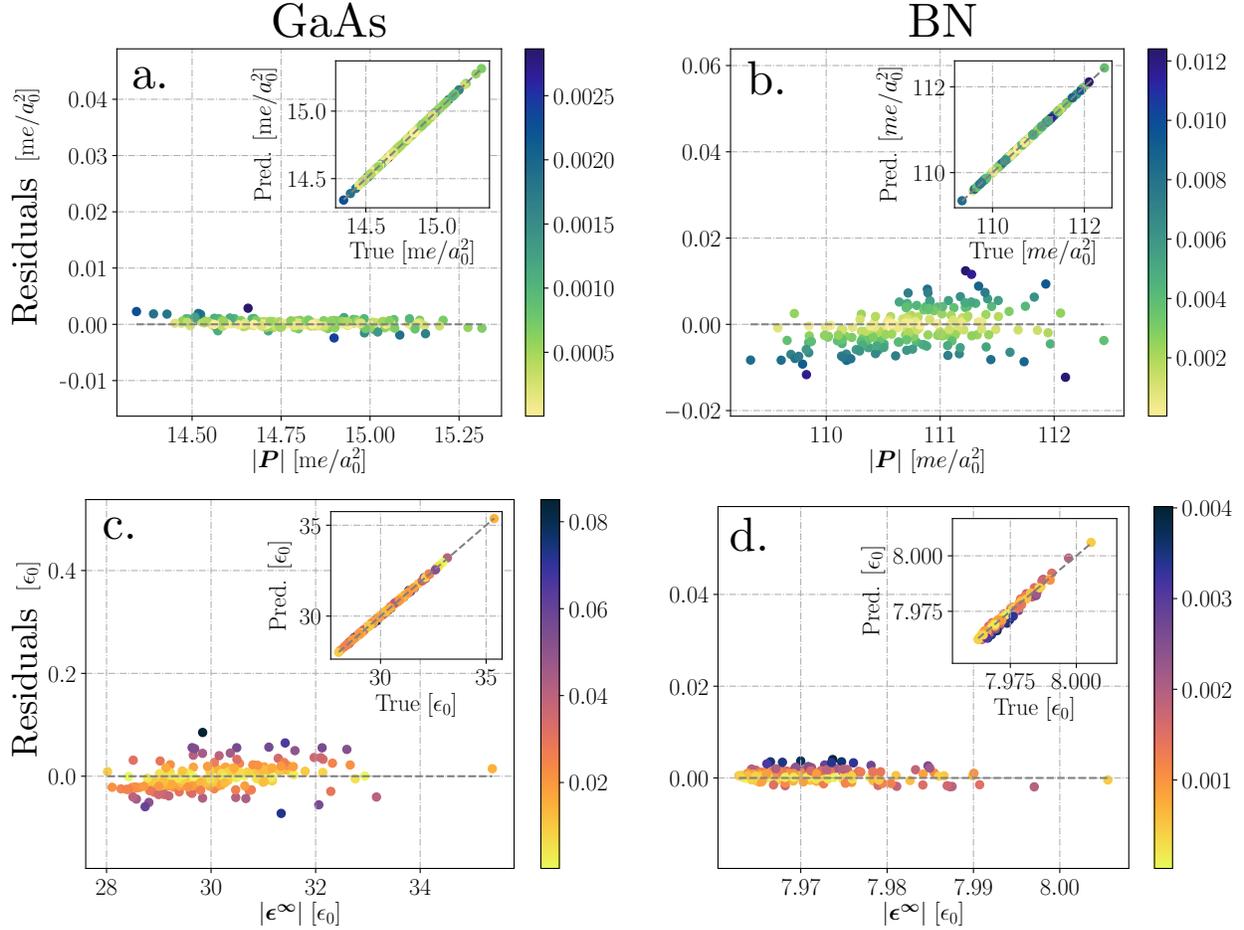}
    \caption{Residuals and true versus predicted (insert) for the absolute values of polarization for (a) GaAs and (b) BN and the static dielectric function for (c) GaAs and (d) BN. See \cref{results} for a description of the computation of the absolute values.}
    \label{fig_2}
\end{figure}

\subsection{Deep Learning}
\label{deep_learning}
As depicted in \cref{fig_1}, \textsc{radnet} utilizes deep neural networks (DNNs) to calculate atomic contributions to the polarization and static dielectric tensor. Similar to Ref. \cite{grisafi2018symmetry}, we write the vector or tensor quantity $\bm{Y}$ as
\begin{equation}
    \bm{Y} = \sum_{m=1}^{N_{\text{ion}}}\bm{Y}_m.
\end{equation}
The decomposition of the desired quantity into atomic contributions allows for an extensive DNN framework \cite{mills2019extensive}. For the polarization, $\bm{Y}$ is a vector with 3 components. For the static dielectric tensor, $\bm{Y}$  is a vector with 6 components. These 6 components correspond to the non-equivalent elements of the tensor. Each component was independently normalized such that $Y_i\in[0,1]$. For the input into the networks, we first construct a real-space representation by summing over Gaussian functions
\begin{equation}
\label{rep}
    \mathcal{R}(\bm{r})=\sum_{m=1}^{N_{\text{ion}}}Z_m \exp(-(\bm{r} - \bm{r}_m)^2 / 2).
\end{equation}
This representation has had success in the past for machine-learned potential energy surfaces \cite{bartok2010gaussian, brockherde2017bypassing, ryczko2018convolutional, mills2019extensive}. The 3-dimensional grid on which \cref{rep} is evaluated must be chosen such that the evaluation time of $\mathcal{R}$ is minimal but the information is not hampered by a poor spatial resolution. In our case, we halved the grid sizes used in the DFT calculations. For GaAs the grid size used was $24\times24\times24$, and for BN the grid size used was $15\times15\times15$.  We then locate the voxels that correspond to the atomic positions and take 3-dimensional slices of the function $\mathcal{R}(\bm{r})$, centered on atomic positions. The dimensions of the slice are defined by a specified cut-off radius, which is used to define a spherical field of view, as shown in \cref{fig_1}a. We experimented with the cut-off radius when predicting $\bm{\epsilon}^{\infty}$ of GaAs and found that a value of 1.852~\AA~(3.5 Bohr) yielded the lowest error on the validation set. We used this cut-off radius for all subsequent calculations. The spherical field of view is achieved by applying a spherical cut-off function, where we considered both a hard cut-off and a smooth one (the complimentary error function). We compared the performance of a model predicting $\bm{\epsilon}^{\infty}$ for GaAs with each cut-off function. We found that both functions performed similarly and chose to use the smooth one. After applying the spherical cut-off function, the slices were fed into a DNN that outputs the contribution to the desired quantity. For the DNN architecture, we used the same DNN as used in Ref. \cite{ryczko2021orbital}. It is comprised of a set of reducing and non-reducing 3-dimensional convolutional layers, followed by a fully connected layer that outputs the atomic contribution to the quantity of interest. The final output is found by summing over the atomic contributions. When training, we used a batch size of 32, a learning rate of $10^{-5}$, and trained for $10^4$ epochs. The models that had the lowest mean squared errors on the validation sets were used for inference. All models were implemented in PyTorch \cite{paszke2019pytorch}. Datasets and codes can be found here \cite{codes}.

\begin{table}[t]
    \centering
    \begin{tabular}{c|c|c|c|c}
        System & Description & Quantity & MAE & RMSE\\\hline\hline
        & & & & \\
         GaAs & Val. Set & $\bm{P}$ [$e / a_0^2$] & $4.25\times10^{-4}$ & $6.78\times10^{-4}$  \\
         GaAs & Val. Set & $\bm{\epsilon}^{\infty}$ [$\epsilon_0$] & $\bm{1.25\times10^{-2}}$ & \bm{$1.77\times10^{-2}$} \\
         GaAs & SchNet & $\bm{\epsilon}^{\infty}$ [$\epsilon_0$] & 0.38 & 0.51 \\
         GaAs & SchNet-$\alpha$ & $\bm{\epsilon}^{\infty}$ [$\epsilon_0$] & 7.61 & 8.93 \\
         GaAs & SA-GPR & $\bm{\epsilon}^{\infty}$ [$\epsilon_0$] & 1.03 & 1.23 \\
         & & & & \\
         GaAs & Equil. & $\bm{P}$ [$e / a_0^2$] & $1.19\times10^{-6}$ & $1.32\times10^{-6}$\\
         GaAs & Equil. & $\bm{\epsilon}^{\infty}$ [$\epsilon_0$] & 0.65 & 0.72 \\
         GaAs & Equil. & $\bm{Z}^*$ [$e$] & $1.04\times10^{-2}$ & $1.43\times10^{-2}$\\ 
         GaAs & Equil. & $\bm{\alpha}$ [$\epsilon_0 / a_0^3$] & $0.30$  & $0.46$\\ 
         GaAs & Equil. & LO-TO [cm${}^{-1}$] & $0.33$ & - \\
         & & & & \\
         BN & Val. Set & $\bm{P}$ [$e / a_0^2$] &$3.85\times10^{-3}$ & $5.18\times10^{-3}$   \\
         BN & Val. Set & $\bm{\epsilon}^{\infty}$ [$\epsilon_0$] & $5.19\times10^{-3}$ & $7.32\times10^{-3}$ \\
         BN & SchNet & $\bm{\epsilon}^{\infty}$ [$\epsilon_0$] & $1.11\times10^{-2}$ & $1.53\times10^{-2}$ \\
         BN & SchNet-$\alpha$ & $\bm{\epsilon}^{\infty}$ [$\epsilon_0$] & 1.85 & 2.17 \\
         BN & SA-GPR & $\bm{\epsilon}^{\infty}$ [$\epsilon_0$] & $\bm{1.38\times10^{-3}}$ & $\bm{2.18\times10^{-3}}$ \\
         & & & & \\
         BN & Equil. & $\bm{P}$ [$e / a_0^2$] & $1.68\times10^{-6}$ & $1.91\times10^{-6}$ \\
         BN & Equil. & $\bm{\epsilon}^{\infty}$ [$\epsilon_0$] & $2.04\times10^{-4}$ & $2.18\times10^{-4}$ \\
         BN & Equil. & $\bm{Z}^*$ [$e$] & $5.65\times10^{-3}$  & $6.52\times10^{-3}$\\ 
         BN & Equil. & $\bm{\alpha}$ [$\epsilon_0 / a_0^3$] & $2.80\times10^{-4}$  & $4.20\times10^{-4}$\\ 
         BN & Equil. & LO-TO [cm${}^{-1}$]& 2.14 & - 

    \end{tabular}
    \caption{Mean absolute errors (MAEs) and root mean absolute errors (RMSEs) for GaAs and BN for different quantities (units in square brackets) in different settings. The description ``Val. Set" means validation set and ``Equil." means equilibrium structure. SchNet, SchNet-$\alpha$, and symmetry-adapted Gaussian process regression (SA-GPR) values are for the validation set. SchNet-$\alpha$ signifies that the polarizability(dielectric) output module was used. Best of the model comparisons are indicated in bold.  }
    \label{table_1}
\end{table}

\section{Results}
\label{results}
We now discuss results of the machine learning models for the computation of the polarization $\bm{P}$ and static dielectric tensor $\bm{\epsilon}^{\infty}$. In \cref{fig_2}, we show the true versus predicted values for $|\bm{P}|$ and $|\bm{\epsilon}^{\infty}|$ for the different systems. For $\bm{P}$, we compute the absolute value via
\begin{equation}
    |\bm{P}| = \sqrt{P_1^2 + P_2^2 + P_3^2}
\end{equation}
and for $\bm{\epsilon}^{\infty}$ we use
\begin{equation}
    |\bm{\epsilon}^{\infty}|=\sqrt{(\epsilon_{11}^{\infty})^2 + (\epsilon_{12}^{\infty})^2 + (\epsilon_{13}^{\infty})^2 + (\epsilon_{22}^{\infty})^2 + (\epsilon_{23}^{\infty})^2 + (\epsilon_{33}^{\infty})^2}.
\end{equation}
In all 4 plots, we see excellent agreement between the true and predicted values. \textsc{radnet} can accurately make predictions despite the changes in the ranges of values of $\bm{P}$ and $\bm{\epsilon}^{\infty}$. In the case of GaAs, the range of $|\bm{\epsilon}^{\infty}|$ is greater than $|\bm{P}|$ ($\approx8$x), whereas for BN we observe the opposite behavior ($\approx80$x). For BN, $\bm{\epsilon}^{\infty}$ varies minimally with the random perturbations of the atomic positions. When comparing the errors in \cref{table_1} for $\bm{P}$ for GaAs and BN, we find that errors are of similar magnitude. Errors in \cref{table_1} for $\bm{\epsilon}^{\infty}$ for GaAs and BN show that BN errors are an order of magnitude smaller than GaAs. However, if we consider the error relative to the range of values shown in \cref{fig_2}, we find that the relative error for GaAs is two orders of magnitude smaller. The quantities are driven both by the model error as well as the range of values being predicted.\\

To compare our methodology with atom-centered approaches, we consider SchNet \cite{schutt2018schnet} and symmetry adapted Gaussian process regression (SA-GPR) \cite{grisafi2018symmetry}. Errors associated with validation sets of these models are reported in \cref{table_1}. In the case of SchNet, we used an output module meant for energy predictions, which we refer to as SchNet, and an output module specifically designed for molecular polarizabilities (dielectric function), which we refer to as SchNet-$\alpha$. In all cases, we used the default parameters included with these packages and radial cut-offs of 5 \AA. We found that SchNet outperforms SchNet-$\alpha$ for both GaAs and BN. In the case of GaAs, SchNet is the best performing alternative to \textsc{radnet}. In the case of BN, SA-GPR outperforms both SchNet and $\textsc{radnet}$. SchNet is an order of magnitude worse and \textsc{radnet} is on-par with SA-GPR. We find the performance of \textsc{radnet} is better or on-par with pre-existing methods. However, an extensive hyperparameter search was not performed for the pre-existing methods. \\

When predicting $\bm{P}$ and $\bm{\epsilon}^{\infty}$ for the equilibrium structures, we find that the errors are less than the validation sets in all cases except when predicting $\bm{\epsilon}^{\infty}$ for GaAs. For both GaAs and BN, $\bm{\epsilon}^{\infty}$ for the equilibrium structure is isotropic. Off-diagonal elements are only non-zero when symmetry is broken. This is true for all training and validation samples in the BN dataset. As discussed in \cref{methods}, for the  GaAs dataset, an additional 200 calculations were included where the $x$-coordinates of the atoms were held fixed. This resulted in more configurations with zero off-diagonal elements. This effect is also true for BN, but the magnitude of the off-diagonal elements is less than those of GaAs. Careful consideration of the desired quantities must be done when constructing an optimal training set.\\

\begin{figure}[t]
    \centering
    \includegraphics[width=0.5\linewidth]{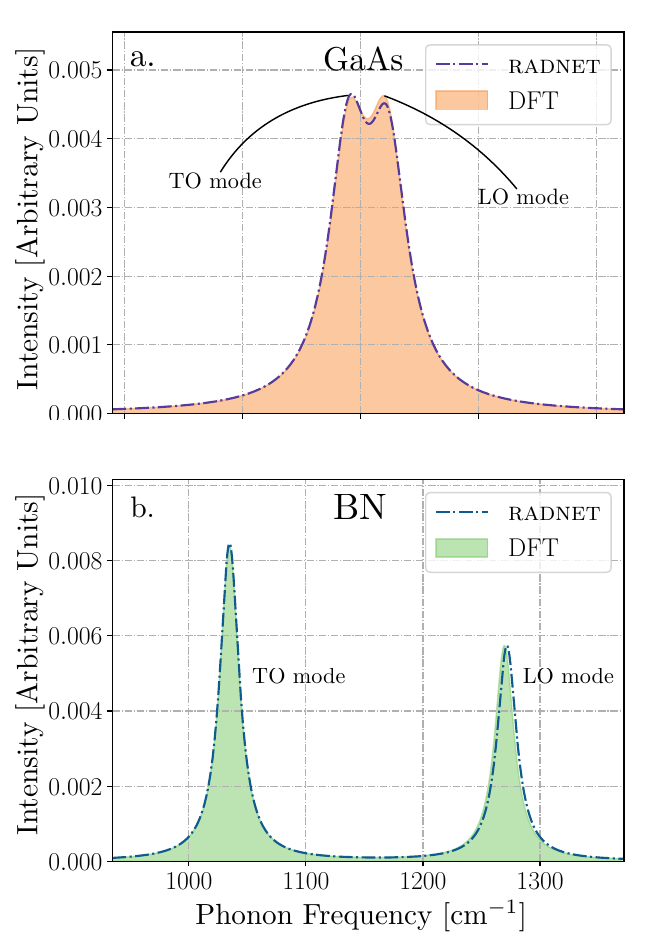}
    \caption{Raman spectra calculated via DFT (shaded regions) and via ML (dashed lines) for (a) GaAs and (b) BN. }
    \label{fig:raman}
\end{figure}

We now discuss the Born-effective charges, $\bm{Z}^*$. As mentioned in \cref{methods}, we compute $\bm{Z}^*$ with finite differences with $\Delta=0.026$ \AA (0.05 Bohr). We also enforce charge neutrality of $\bm{Z}^*$ by following the sum rules outlined in Ref. \cite{gonze1997dynamical}. Looking to \cref{table_1}, we note that for both GaAs and BN, the errors for $\bm{Z}^*$ are 3-4 orders of magnitude larger than their respective $\bm{P}$ errors. There are two contributions to this error. The first is the difference in ranges for these quantities. The range-driven error causes the magnitude of $\bm{Z}^*$ to be 2-3 orders of magnitude larger than $\bm{P}$ which accounts for the majority of the error differences. The remainder of the error can be attributed to errors associated with the machine learning model. In some cases, over-fitting can be reduced with the use of regularization techniques as was shown in a previous report for phonon calculations \cite{malenfant2021}. However, in our case, we found that the use of $L2$ regularization did not improve the derivative predictions and therefore conclude that over-fitting is not a factor. The remainder of the error could be improved by an improved DNN architecture. However, the current accuracy of the predictions is satisfactory and its use in subsequent quantities yielded good agreement with conventional approaches.  \\

We now discuss LO-TO splitting. In \cref{loto}, the TO phonon energy is needed to calculate the splitting, which yields the LO energy. We used the DFT phonon energies from the equilibrium structures along with the machine-learned quantities to calculate the splitting. We note that one can also use various machine learning techniques to calculate the phonon energies if one has a differentiable model which explicitly depends on atomic coordinates. We find that errors of the LO modes for both GaAs and BN are less than 0.2\%. In a past report \cite{janssen2010electron}, phonon energies calculated via DFT with different hybrid functionals were found to have maximum errors of $\approx 5$\% when compared to experiment. The predictions we report are well within this range. Despite having a large error in the prediction of $\bm{\epsilon}^{\infty}$ for the equilibrium structure of GaAs, we achieve excellent agreement when comparing the LO-TO splitting frequencies. This is due to the choice of the vector $\bm{q}$ and the form of \cref{loto}. The choice of $\bm{q}$-vector when calculating the LO-TO splitting is arbitrary, but reasonable choices are the Cartesian unit vectors.  This particular choice eliminates the error from off-diagonal elements of $\bm{\epsilon}^{\infty}$ due to the denominator in \cref{loto}. With $\bm{q}=\hat{x}$ Bohr${}^{-1}$, the first diagonal term is selected. With $\bm{q}=\hat{y}$ Bohr${}^{-1}$ the second diagonal term, and with $\bm{q}=\hat{z}$ Bohr${}^{-1}$ the third diagonal term. If one is only interested in computing the Raman spectra for these choices of $\bm{q}$-vectors then the off-diagonal terms can be omitted entirely from the datasets. The value of the LO-TO splitting frequency, as reported in \cref{table_1}, is averaged over the 3 unit directions. \\

We now discuss computation of the Raman spectra. Using finite differences, we compute the derivatives with respect to atomic displacement, thereby yielding the Raman tensors, $\bm{\alpha}$ shown in \cref{raman-tensors}. We used the DFT phonon eigenvectors from the equilibrium structures along with the machine-learned quantities to calculate $\bm{\alpha}$. As noted previously, if one has a fully differentiable machine learning model that explicitly depends on atomic coordinates, it is possible to calculate the phonon eigenvectors using automatic differentiation. In \cref{table_1}, we report errors of $\bm{\alpha}$. For both GaAs and BN, we find that these errors are on-par with their respective $\bm{\epsilon}^{\infty}$ counterparts. Using the phonon energies calculated via DFT, along with the LO-TO splitting and $\bm{\alpha}$ values calculated via machine learning, we plot the Raman spectra for GaAs and BN in \cref{fig:raman}. Agreement of the LO-TO splitting is observed once again for both GaAs and BN when comparing the LO modes. In addition, the peak heights are also in good agreement due to the accurate predictions of $\bm{\alpha}$. This is promising for future studies involving larger-scale systems. In addition to the excellent agreement, the computational speed-up from our inference script in comparison to a DFT calculation of $\bm{\epsilon}^{\infty}$ was $\approx$250 times faster. Inference timing includes checkpoint loading, input generation, model inference, and derivative calculations (which were all done on 32 CPU cores). This accuracy and computational speed-up show promise to study the Raman response of large-scale systems. The relative height changes of peaks with respect to change in defect concentration have been done experimentally \cite{mignuzzi2015effect} for MoS${}_2$, and a future application of \textsc{radnet} will be studying the effects of defects in the Raman response of solids. To do so, two limitations of \textsc{radnet} must be addressed. The first limitation is the used of finite differences to evaluate derivatives. In order to perform inference on larger-scale systems, automatic differentation must be integrated. The second limitation is the implementation of the input function generation $\mathcal{R}(\rr)$, which is currently evaluated on the entire grid. Due to the locality of Gaussian functions, a radial cut-off can also be applied when evaluating the Gaussian on a numerical grid to improve the evaluation time. Therefore, an efficient, parallel implementation for the input function generation, similar to what was used in Ref. \cite{mills2019extensive}, must also be implemented.  \\

\section{Conclusion} 
\label{conclusion}
We introduce  a deep learning approach, called the \textbf{r}eal-space \textbf{a}tomic \textbf{d}ecomposition \textbf{net}work, or \textsc{radnet}, for predicting the polarization and the static dielectric tensor in solids.  We find excellent agreement when comparing predictions to true values and the relative error for both GaAs and BN are of similar magnitude. We also compare our results to other approaches when considering the dielectric function, and find that \textsc{radnet} has the best result for GaAs, and is on-par with symmetry-adapted Gaussian process regression \cite{grisafi2018symmetry} for BN. After performing inference to obtain the polarization and infinite dielectric tensor, we then used these predictions to calculate Born-effective charges, the LO-TO splitting, and Raman tensors. We find good agreement when comparing the machine-learned quantities to DFT, and used the LO-TO splitting frequencies and Raman tensors along with DFT phonon energies and eigenvectors to compute the Raman spectra. The LO-TO splitting agreement is again confirmed when comparing the Raman spectra computed with the machine-learned quantities and DFT. Shifts of the LO modes as well as the peak heights are in exceptional agreement. Future work will include studying condensed matter systems with defects to examine the effects on the Raman spectra of realistic defect concentrations.

\section{Acknowledgements}
The authors acknowledge fruitful discussions with Arnab Majumdar. The authors also acknowledge Compute Canada, the Artificial Intelligence for Design (AI4D) program at the National Research Council, and the Vector Institute for Artificial Intelligence for computational resources. The authors also acknowledges the National Sciences and Engineering Council of Canada. KR acknowledges the Vector Institute for Artificial Intelligence for funding.

\bibliography{refs}
\end{document}